\documentclass[a4paper,11pt]{article}

\usepackage{jheppub} % for details on the use of the package, please
\usepackage[T1]{fontenc} % if needed
\usepackage{latexsym,amsfonts,amssymb,amsthm,amsmath}
\usepackage{physics}
\usepackage{xcolor}
\usepackage{simpler-wick}
\usepackage{comment}
\usepackage{subfig}
\usepackage{bbold}
\usepackage[normalem]{ulem}

\usepackage{multicol}

\usepackage{tikz}
\usepackage[compat=1.1.0]{tikz-feynman}
\usetikzlibrary{external}
\immediate\write18{mkdir -p pgf-img}
\title{\boldmath Open-Source Numerical Solver for Neutrino Collective Effects - I: Isotropic Neutrino Gas}

\author[a,1]{P. Dedin Neto,\note{Corresponding author.}}

\affiliation[a]{University of Campinas -- UNICAMP,\\Campinas, 13083-859, Brazil}

\emailAdd{dedin@ifi.unicamp.br}

\abstract{ In this paper, we introduce a new open-source code to find numerical solutions for the neutrino evolution considering neutrino-neutrino interactions, which result in the so-called collective effects. We first describe the theoretical background of this type of evolution, explaining the polarization vector formalism in which we implement our numerical code. We then show the results for different neutrino systems. In this first paper, we focus on the case of an isotropic neutrino gas, exploring a mono-energetic scenario and one with a spectral distribution. The resulting code of this work is available at \url{https://github.com/pedrodedin/Neutrino-Collective-Effects}.}

\begin{document} 
\maketitle
\flushbottom

\section{Introduction}
\label{sec:intro}

When considering environments with a high density of neutrinos, such as the case of a supernova and the early universe, we have to consider the potential due to neutrino-neutrino forward scattering. However, the inclusion of neutrino-neutrino interactions makes the evolution equation nonlinear and complicated to solve in many cases. For supernova neutrinos, it became extremely challenging due to many complications (such as non-uniform emission, energy and flavor-dependent neutrinospheres, among others) and is still an unsolved problem. Despite the difficulty, there are many works in the literature trying to solve this problem. Some of these works show analytical solutions for simpler neutrino systems \citep{Hannestad2006-zw,Duan:2005cp}, such as mono-energetic and isotropic neutrino gas. Others take the numerical approach and consider more complex systems \citep{Duan2006-dp} \footnote{See \citep{Duan2010-dq,Mirizzi:2015eza} for reviews on the topic.}. Even so, we still do not have a clear picture of the neutrino evolution, given that new phenomena emerge when other features are considered, such as the case of fast oscillations due to ELN (electron lepton number) crossing in the angular distribution \citep{Tamborra:2020cul,Sawyer:2008zs}.  

In this context, this paper aims to describe a new open-source code for numerical solutions to the neutrino evolution when considering neutrino-neutrino interactions. In this first version, we explore the case of an isotropic neutrino gas. This system is relatively simple and has already been explored in the literature. However, when talking about collective effects, there is a lack of open-source codes from the papers that present numerical solutions, which makes it difficult to reproduce and analyze the results. Therefore, our work is an effort to try to change this picture, offering a code with which anyone can reproduce our results and make further improvements. Moreover, we intend to continue to improve this code, exploring increasingly supernova-like scenarios in future works.

This document is organized as follows. In section \ref{sec:Evolution_Equation}, we describe the theoretical background of the neutrino evolution equation, in which we consider a free-streaming neutrino influenced by forward scattering potentials due to neutrino-electron and neutrino-neutrino interactions. In the same section, we describe the framework of polarization vectors that we use both in the code implementation and to analyze the results. In section \ref{sec:Numerical_Implementation}, we describe how the numerical implementation is done, listing the important steps of the code and the libraries used. In section \ref{sec:Numerical_Solutions} we show the numerical results for the different scenarios considered up to this moment. Finally, in section \ref{sec:Conclusions} we make our conclusions.

\section{The Evolution Equation}
\label{sec:Evolution_Equation}
\subsection{Evolution Equation - Free-streaming Neutrino}

\begin{figure}[hbt!]
 \centering
     \subfloat[][$\nu-e$ scattering]{
         \label{fig:H_nu-e}
        \feynmandiagram [vertical=a to b] {
         i1 [particle=\(\nu_{e}(p_{1})\)] -- [fermion] a -- [fermion] f1 [particle=\(e^- (p_{2})\)],
         a -- [boson, edge label=\(W^{+}\)] b,
         b -- [fermion] i2 [particle=\(\nu_e(p_{1})\)],
         f2 [particle=\(e^- (p_{2})\)] -- [fermion] b,
        };
    }
    \subfloat[][$\nu-\nu$  scattering]{
        \label{fig:H_nu-nu}
        \feynmandiagram [vertical=a to b] {
          i1 [particle=\(\nu_{\alpha}(p_{1})\)] -- [fermion] a -- [fermion] f1 [particle=\(\nu_{\alpha}(p_{1})\)],
          a -- [boson, edge label=\(Z^{0}\)] b,
          b -- [fermion] i2 [particle=\(\nu_{\beta}(p_{2})\)],
          f2 [particle=\(\nu_{\beta}(p_{2})\)] -- [fermion] b,
        };
        \begin{tikzpicture}
        \begin{feynman}
            \diagram [vertical=a to b] {
            i1 [particle=\(\nu_{\alpha}(p_{1})\)]
            -- [fermion] a
            -- [draw=none] f1 [particle=\(\nu_{\beta}(p_{1})\)],
            a -- [photon, edge label'=\(Z^{0}\)] b,
            
            b -- [draw=none] f2 [particle=\(\nu_{\alpha}(p_{2})\)],
            
            i2 [particle=\(\nu_{\beta}(p_{2})\)] -- [fermion] b,
            };
            
            \diagram* {
            (a) -- [fermion] (f2),
            (b) -- [fermion] (f1),
            };
        \end{feynman}
        \end{tikzpicture}
    }
\caption{Feynman diagrams for the neutrino forward scattering in electrons and neutrinos.}
\label{fig:nu-nu_diagonal}
\end{figure}
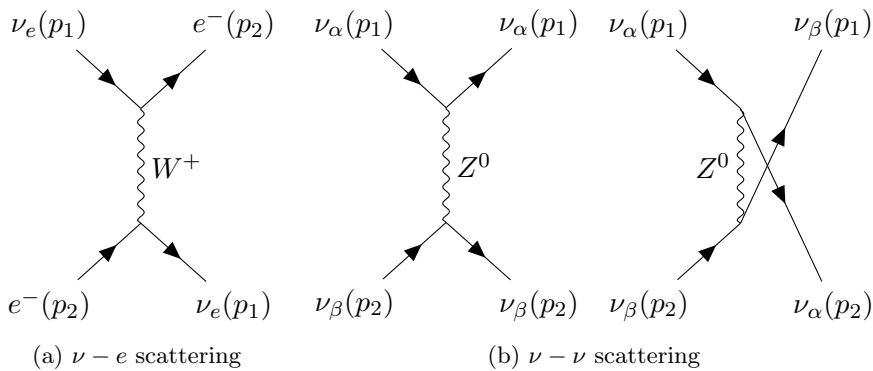

For any quantum system, its evolution is given by the Hamiltonian. In the Schrodinger picture, the evolution of a quantum state $\ket{\psi}$ is given by the following equation:
\begin{equation}
\label{eq:evolution}
    i\frac{d}{dt} \ket{\psi (t)} =   H \ket{\psi (t)} = (H_0 + H_{int}) \ket{\psi (t)},
\end{equation}
where the Hamiltonian can be divided into a free $H_0$ and an interacting part $H_{int}$ \footnote{It is worth mentioning that we consider the time $t$ and propagate distance $r$ as interchangeable throughout the paper, given that the neutrino is ultra-relativistic ($E\gg m_\nu$) in the cases interesting for us.}. For a free-streaming neutrino, i.e., a neutrino that is not scattered away from its trajectory from the source to the detector, the interaction part of the Hamiltonian is composed of forward-scattering potentials. That is the case of solar neutrinos and supernova neutrinos free-streaming above the neutrinosphere. For supernova neutrinos, the relevant potentials are the ones due to scattering in ordinary matter (electrons, protons, and neutrons) and neutrinos themselves. For interactions with ordinary matter, if we consider only active neutrinos, the potential resulting from scattering in protons and neutrons is irrelevant, as it only contributes to an overall phase. In this case, the only contribution that matters is the one due to the neutrino-electron charged current (CC) interaction (Figure \ref{fig:H_nu-e}), for which the potential in the flavor basis is given by
\begin{equation}
     H_{\nu e} = \sqrt{2} G_F n_e \text{diag} \left[1,0,0\right]
\end{equation}
considering three families of neutrinos, where $G_F$ is the Fermi constant and $n_e$ is the local electron density \citep{wolfenstein1978neutrino}. For the neutrino-neutrino interactions (Figure \ref{fig:H_nu-nu}), the potential is given by the following expression \citep{Neto:2021hhl,sigl1993general}:
\begin{equation}
     H_{\nu \nu} = \sum_{j=1}^N
    \sqrt{2} G_F (1-\cos \theta_{ij}) [\rho_{j}(t)-\overline{\rho}_{j}(t)],
\end{equation}
where $\theta_{ij}$ is the angle between the momentum $\vec{p}_i$ of the evolved neutrino and the momentum $\vec{p}_j$ of the $j$-th neutrino or antineutrino interacting with the evolved one, such that their density matrices are given respectively by $\rho_{j}(t) = \ket{\nu_j(t)}\bra{\nu_j(t)}$ and $\overline\rho_{j}(t) = \ket{\overline{\nu}_j(t)}\bra{\overline{\nu}_j(t)}$. The indices $i$ and $j$ may as well be exchanged by a momentum $\vec{p}$ index, this just depends on which way we want to describe our neutrinos (individually or by an ensemble of a given momentum). It is also worth mentioning that the electrons and neutrinos interacting with $\ket{\psi (t)}$ are the ones that have a wave-packet overlap with it, so that the quantities in the equations above, such as electron and neutrino densities, correspond to their local value in the neutrino trajectory \footnote{In this sense, it is useful to imagine the neutrino wave function spatial distribution $\psi (\vec{r},t)$ when looking to the equation (\ref{eq:evolution}), rather than just an abstract state $\ket{\psi}$}. A good discussion about the microscopic nature of the neutrino evolution can be found on \citep{Akhmedov:2020vua}.

\subsection{Polarization Vector Formalism}

Given the evolution equation (\ref{eq:evolution}), we will now translate it into the polarization vector formalism, the one in which we implement our code. First, we can describe a quantum state $\ket{\psi(t)}$ using the density matrix, which is defined as follows:
\begin{equation}
    \rho(t) = \ket{\psi(t)}\bra{\psi(t)}.
\end{equation}
We can also define the density matrix for a statistical ensemble of states (incoherent mixture), where each state $\ket{\psi_i(t)}$ have a statistical weight $p_i$. The last case represents what is called a \textit{mixed state}, and the former case is called a \textit{pure state}. Using equation (\ref{eq:evolution}), we can write the evolution equation for the density matrix as
\begin{equation}
    i\frac{d\rho}{dt} = [H , \rho]
\end{equation}
If we consider a scenario of two families of neutrinos, we can decompose the matrices above in Pauli matrices. Then, we can use the coefficients of this expansion as components of three-dimensional vectors that represent the matrices in the following way:
\begin{equation}
     H = -\frac{1}{2}\vec{\sigma} \cdot \vec{H} \text{,} \;\;\;\;\; \rho= \frac{1}{2} \mathbf{1} + \frac{1}{2} \vec{\sigma} \cdot \vec{P}\text{,} \;\;\;\;\; \vec{\sigma}=(\sigma_1,\sigma_2,\sigma_3).
\end{equation}
Here we ignore the Hamiltonian component proportional to the identity, which is irrelevant for the evolution. Using this vector notation, the evolution equation becomes a precession one 
\begin{equation}
\label{eq:Precession}
      \frac{d}{dt} \vec{P}(t)= \vec{P}(t)\times \vec{H} \;\;\;\leftrightarrow  \;\;\; i\frac{d}{dt} P_i(t) = \frac{i}{2} P_i(t) H_j [\sigma_i,\sigma_j].
\end{equation}
That is the differential equation that we implement in our code. Note that the coefficients of expansion in the Pauli matrices $\sigma_i$ depend on the basis that we are working on, where the 3rd component $P_3$ will give us the content of eigenstates in the state $\ket{\psi(t)}$ for a given basis\footnote{The Pauli matrix $\sigma_3$ is the only one with non-zero diagonal components.}. Therefore, we just need to project $\vec P$ in the third component to get the probability of detecting a specific eigenstate. The main motivations to work in this formalism, as we will see in the results, are the nice pictorial visualization of the neutrino evolution and the number of classical and well-known equivalent systems.

% so that in the flavor basis the probability of detecting the neutrino in a flavor $\alpha$ is given by

% \begin{equation}
%         P_\alpha(t) =  
%         \left\{ \begin{matrix}
%          P_e(t)= \frac{1}{2}(1+P_3^{F}(t))\\ 
%          P_x(t)= \frac{1}{2}(1-P_3^{F}(t))
%         \end{matrix}\right.
% \end{equation}

\subsection{Polarization Vectors and Forward-scattering Interactions ($H_{\nu e}$ and $H_{\nu\nu}$)}

Let us now consider the specific case of free-streaming neutrinos where the non-interacting component of the Hamiltonian is given by the vacuum Hamiltonian $H_V$ and the interacting part is given by the potentials due to forward scattering in electrons $H_{\nu e}$ and in neutrinos themselves $H_{\nu \nu}$. In this case, we can write the evolution equation for a given neutrino (labeled by the index $i$) as
\begin{equation}
\begin{split}
      i \frac{d}{dt} \rho_i= [\omega H_{V}+ \lambda H_{\nu e}+\mu H_{\nu\nu,i}, \rho_i], \\
\omega \equiv \frac{\Delta m^2}{2E_i},\;\;\;\; \lambda \equiv  \sqrt{2}G_F n_e ,\;\;\;\; \mu \equiv  \sqrt{2}G_F n_\nu,
\end{split}
\end{equation}
where $\omega$, $\lambda$, and $\mu$ give the characteristic intensity of its respective component in the Hamiltonian. Again, we can translate the matrices into vectors by decomposing them in Pauli matrices such that
\begin{equation}
    H_V = -\frac{1}{2}\vec{\sigma} \cdot \vec{B} \text{,} \;\;\;\;\;   H_{\nu e} = -\frac{1}{2}\vec{\sigma} \cdot \vec{L} \text{,} \;\;\;\;\; \rho= \frac{1}{2} \mathbf{1} + \frac{1}{2} \vec{\sigma} \cdot \vec{P}\text{.}
\end{equation}
With this decomposition, the evolution of the i-th neutrino and antineutrino polarization vectors ($\vec{P}_{\nu_i}$ and $\vec{P}_{\overline{\nu}_i}$) interacting with j (anti)neutrinos becomes
\begin{equation}
\label{eq:Precession_Forward}
\begin{split}
    \frac{d}{dt}\vec{P}_{\nu_i} = \vec{P}_{\nu_i} \times \left[ \omega \vec{B} + \lambda \vec{L} - \mu \sum_j (1-\cos{\theta_{ij}}) (\frac{n_{\nu,j}}{n_\nu} \vec{P}_{\nu,j} - \frac{n_{\overline\nu,j}}{n_\nu}\vec{P}_{\overline{\nu},j}) \right], \\
    \frac{d}{dt}\vec{P}_{\overline{\nu}_i} = \vec{P}_{\overline{\nu}_i} \times \left[ - \omega \vec{B} + \lambda \vec{L}- \mu \sum_j (1-\cos{\theta_{ij}}) (\frac{n_{\nu,j}}{n_\nu} \vec{P}_{\nu,j} - \frac{n_{\overline\nu,j}}{n_\nu}\vec{P}_{\overline{\nu},j}) \right].
\end{split}
\end{equation}
Note that it is possible to redefine $\vec{P}$ as to be normalized by $n_\nu$. One just needs to be careful when translating it back to the density matrix or a conversion or survival probability. We can also turn this equation more compact using the neutrino flavor isospin (NFIS) notation $\vec{s}$ \citep{Duan:2005cp}. Even so, we will not use it in this work, keeping the notation of polarization vectors evolved by equation (\ref{eq:Precession_Forward}). In the next section, we describe how we implement this equation in our numerical code.

% \begin{equation}
%     \frac{d}{dt}\vec{s}_{i} = \vec{s}_{i} \times \left[ \omega \vec{B} + \lambda \vec{L} + \mu \sum_j \mu_{ij} \vec{s}_{j}\right] = \vec{s}_{i} \times \left[ \omega \vec{B} + \lambda \vec{L} + \mu \vec{S} \right],\;\;\;\;  s_\nu \equiv \frac{\vec{P}_\nu}{2} \;\;\;\;s_{\overline{\nu}} \equiv - \frac{\vec{P}_{\overline{\nu}}}{2}.
% \end{equation}

% However, in most of this work we will use the polarization vector $\vec{P}$ notation.

\section{Numerical Implementation}
\label{sec:Numerical_Implementation}

Our code is implemented in Python and can be found in its respective repositories on GitHub (\url{https://github.com/pedrodedin/Neutrino-Collective-Effects}) and Zenodo \citep{pedro_dedin_neto_2022_5964385}, along with instructions on how to use it. For the numerical solver, we use the scipy.integrate library \citep{2020SciPy-NMeth} through its \textit{odeint} routine, which solves a system of first-order ODEs (Ordinary Differential Equations) \footnote{It is possible to solve an n-order ODE passing it as a system of n first-order ODEs.}. To solve a system of $n$ ODEs, the solver requires a Python function that computes the derivative $dy_i/dt=f_i (y,t,\theta)$ of the $n$ variables $y_i(t)$. It also required an $n$-dimensional array $y_0$ with the initial conditions and a time array $t$ with the points in which $y_i(t)$ will be solved.

In our case, the number of differential equations depends on how many variables $\vec{P}_{\nu_i}$ depend on, that is, on how many different polarization vectors we divide the neutrinos. In the simplest case, we have only one polarization vector for each flavor of neutrinos and antineutrinos, resulting in four vectors to be evolved. Considering two families of neutrinos, each vector has three components, totalizing $n=12$ differential equations. In a more general case, we may write the number of equations as $n=2\times n_{f} \times n_{dim} \times n_{others}$, where $n_f$ is the number of neutrino flavors, $n_{dim}$ the dimension of the space where the polarization vectors live\footnote{In fact, the dimension $n_{dim}$ of this space can be easily computed from the number of flavors $n_f$, given that the complex $n_f\times n_f$ matrices are decomposed in the generators of the $SU(n_f)$ group, apart from the identity.}, and $n_{others}$ the other possible degrees in which $\vec{P}_{\nu_i}$ or $\vec{P}_{\overline{\nu}_i}$ could depend. For example, in the case of spectral distribution, $n_{others}$ represents the number of energy bins. 

With respect to the time sampling step $\Delta t$ used in the code, it needs to be at least two times smaller than the smallest oscillation period that composes the time variation of $y_i(t)$, as requested by the Nyquist-Shannon sampling theorem. Therefore, a first bound is to consider a $\Delta t$ smaller than the smallest vacuum oscillation period $2\pi/\omega_{max}$, which gives us a good reconstruction for the vacuum oscillatory behavior. The other potentials, such as $\lambda$ and $\mu$, and their variation may also give rise to phenomena with a characteristic time shorter than the vacuum ones. For these effects, instead of using an explicit expression for $\Delta t$ we make empirical adjusts, using a sampling bellow the scale of the dominant potential ($\omega$,  $\lambda$, or $\mu$). It is worth mentioning that the sampling step passed to the ODE solver is only to sample the output, while the integration step used to solve the system of equations is adjusted internally by the solver routine.

%The code can be found in the GitHub or Zenodo repositories \citep{pedro_dedin_neto_2022_5964385}. It is structured in the following Python scripts files:

%\begin{itemize}
%    \item \textbf{Auxiliar\_Functions.py} : Contains auxiliary functions used in the numerical solver, such as matter potential profiles $\lambda (r)$, neutrino-neutrino profiles $\mu (r)$, functions to read the output of the numerical solver, etc.
    
%    \item \textbf{ODE\_Scenario.py} : Contains the numerical ODE solver for each specific scenario (Solar, Isotropic\_Monoenergetic, Isotropic\_Spectrum).
    
%    \item \textbf{Plots.py}: Contains functions for usual plots and animations.
%\end{itemize}

%There is also a Quick Guide Python Jupyter notebook (\textbf{Quick\_Guide.ipynb}) to assist the user in the usage of the code.

\section{Numerical Solutions for Different Scenarios}
\label{sec:Numerical_Solutions}

\subsection{0\textsuperscript{th} Case - Solar Neutrinos (MSW Effect) }

Before diving into the cases with neutrino-neutrino interactions, we will explore the neutrino evolution for the well-known case of solar neutrinos. We first show the analytical solution and then compare it to our numerical one. That serves as a test for the ODE numerical solver and for the framework that we are using.

\subsubsection{Analytical solution}

In the solar case, $\nu_e$ neutrinos are created inside the sun in a high density environment, which generates a matter potential due to neutrino-electron forward scattering. This potential changes the mixing parameters between the flavor states $\ket{\nu_\alpha}$ and the mass eigenstates in matter $\ket{\nu^M_i}$. For two neutrino families, we have in the following effective mixing angle and mass squared difference for a neutrino propagating in matter:
\begin{equation}
\label{eq:eff_mixing_param}
            \tan{2\theta_M} = \frac{\tan{2\theta}}{1-\frac{A_{CC}}{\Delta m^2 \cos{2\theta}}}, \;\;\;\;\; \Delta m^2_M = \sqrt{(\Delta m^2\cos{2\theta}-A_{CC})^2+(\Delta m^2\sin{2\theta})^2},
\end{equation}
\begin{equation}
    A_{CC}\equiv \sqrt{2} G_F n_e E , \;\;\;\;\; A^R_{CC} = \Delta m^2\cos{2\theta},
\end{equation}
where $A^R_{CC}$ defines the resonance condition, in which $\Delta m^2_M$ is minimum and the mixing is maximum ($\theta_M=\pi/4$). As the neutrino escapes the Sun, the electron density $n_e (r)$ changes over its trajectory, leading to a flavor conversion described by the MSW (\textit{Mikheyev}-\textit{Smirnov}-\textit{Wolfenstein}) effect \citep{wolfenstein1978neutrino,mikheev1985resonance}. In summary, the evolution can be \textit{adiabatic} if the electron density varies slowly enough for the neutrino to remain in the same instantaneous mass eigenstate, or n\textit{on-adiabatic} if the variation is fast enough to cause a transition between the instantaneous eigenstates near the resonance region. For the solar neutrinos, we have an adiabatic evolution, so that a neutrino created in a specific superposition of effective mass eigenstates will stay in this same superposition as the electron density decreases to zero. This initial superposition will depend on the potential $A^{(i)}_{CC}$ at the neutrino creation such that for an electron neutrino, we have the following initial state:
\begin{equation}
    \ket{\nu_e (0)} = \cos{2\theta_M^{(i)}}\ket{\nu_1^M} + \sin{2\theta_M^{(i)}} \ket{\nu_2^M} \approx \left\{\begin{matrix}
\ket{\nu_2^M} &, A^{(i)}_{CC} \gg A^R_{CC}\\
\cos{2\theta}\ket{\nu_1} + \sin{2\theta} \ket{\nu_2}  &, A^{(i)}_{CC} \ll A^R_{CC} .
\end{matrix}\right.
\end{equation}
Given that the evolution is adiabatic, to get the survival and conversion probabilities, we just have to replace the mass eigenstates in matter by their respective ones in the vacuum and use the known mixing angle in vacuum $\theta$. Therefore, we can summarize the solar neutrino survival probability as follows:
\begin{equation}
    \langle P_{\nu_e \rightarrow \nu_e} \rangle_L \approx 
    \left\{\begin{matrix}
    \frac{1}{2} - \frac{1}{2}\sin^2{2\theta} &, E<2 \text{MeV} &\text{(Low Energy)}\\ 
    \frac{1}{2} + \frac{1}{2} \cos{2\theta_M^{(i)}}\cos{2\theta} &, E>2 \text{MeV} &\text{(Moderated Energy)}\\ 
    \sin^2{\theta}  &, E \gg 2 \text{MeV} &\text{(High Energy)} ,
    \end{matrix}\right.
\end{equation}
where we consider the probability averaged over propagated length $L$ due to the uncertainty on its exact value.

\subsubsection{Numerical Solution}

\begin{figure}[h]
    \centering
    \includegraphics[trim= 0 0 530 0, clip,scale=0.5]{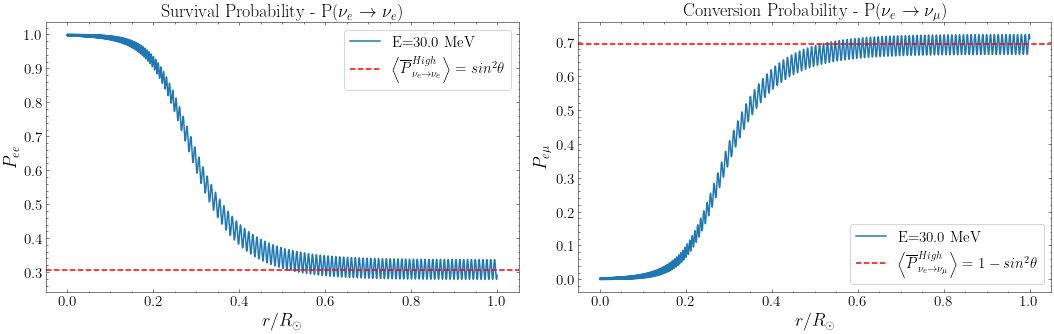}
    \caption{Numerical survival probability for high energy ($E_\nu=30$ MeV) solar neutrinos in blue, and the analytical averaged probability in red.}
    \label{fig:solar_high_E_prob}
\end{figure}

\begin{figure}[h]
    \centering
    \includegraphics[scale=0.5]{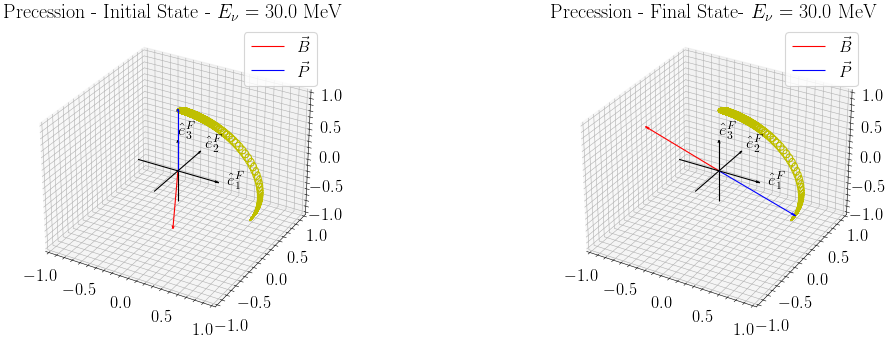}
    \caption{Here we show the polarization vectors $\vec{P}$ (blue with trajectory in yellow) and $\vec{B}$ (red), corresponding to the results shown in figure \ref{fig:solar_high_E_prob}. The left plot is a snapshot of the initial state, while the right one corresponds to the final state outside the Sun (See the animation in \url{https://github.com/pedrodedin/Neutrino-Collective-Effects/blob/main/Figures/Solar_High_Energy.gif}).}
    \label{fig:solar_high_E_vec}
\end{figure}

For the numerical solution, we have adopted the mixing parameters $\Delta m^2_{21} = 7.53 \times 10^{-5} \text{eV}^2$ and $\sin^2{\theta_{12}}=0.307$ \citep{Zyla:2020zbs}. For the electron density profile inside the Sun, we adopted the following parametrization \citep{Bahcall:2000nu}:
\begin{equation}
    n_e(r) = 245 N_A exp(-10.45 r/R_\odot),
\end{equation}
where $r$ is the radial distance from the center of the Sun, $N_A$ is the Avogadro number and $R_\odot$ is the solar radius. The resulting outcome from our numerical solver can be seen in figures \ref{fig:solar_high_E_prob} and \ref{fig:solar_high_E_vec}  for high energy neutrinos. In the figure, we can see the trajectory of the neutrino polarization vector $\vec{P}$ and the resulting survival probability, which is in agreement with the average one from the analytical solution. As expected, in the high-energy case, the electron neutrino is created in the heavier mass eigenstate $\ket{\nu_e} \approx \ket{\nu_2^M}$, which corresponds to a $\vec{P}$  anti-parallel to $\vec{B}$ and remains in the same state, coupled to $\vec{B}$, as the density decreases towards the vacuum. These results show that our code and approach are in good agreement with the expected analytical solutions, which allows us to proceed to the scenarios with neutrino-neutrino interactions.

\subsection{1\textsuperscript{st} Case - Isotropic and Mono-energetic Neutrino Gas}

The first case with neutrino-neutrino interactions that we treat is the one corresponding to an isotropic ($ \left \langle \cos{\theta_{ij}}\right \rangle = 0$), mono-energetic ($\omega_i=\omega$) and uniform neutrino gas composed of neutrinos with opposed neutrino flavor isospin, also known as \textit{bipolar systems}. That is the case of a system composed of neutrinos with opposite polarization vectors or a neutrino-antineutrino system of the same flavor. For simplicity, we will not consider matter effects ($\lambda = 0$). The presence of a uniform matter potential will only change the mixing parameters to those effective in matter, which does not change our conclusions. Regarding the vacuum mixing parameters, we will use the following ones: $\Delta m^2_{31}= 2.5 \times 10^{-3} eV^2$ and $\sin^2{\theta_{13}} = 2.1 \times 10^{-2}$ \cite{Zyla:2020zbs}, which are more representative of a supernova scenario.

\subsubsection{Symmetric System and Constant $\mu$}

\begin{figure}[h]
    \centering
    \includegraphics[scale=0.5]{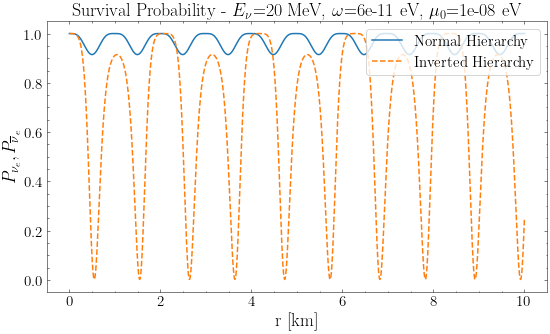}
    \caption{Numerical survival probability for the symmetric and mono-energetic neutrino gas with constant neutrino potential $\mu= 1\times 10^{-8}$ eV and energy $E=20$ MeV.}
    \label{fig:SN_Bipolar_fixed_E_and_mu_prob}
\end{figure}

\begin{figure}[h]
    \centering
    \includegraphics[width=0.4\textwidth]{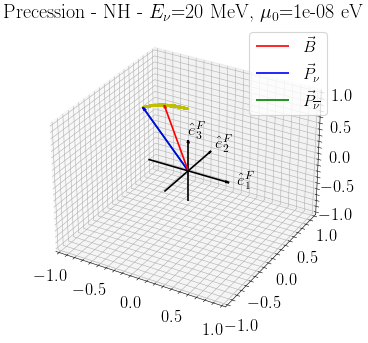}
    \includegraphics[width=0.4\textwidth]{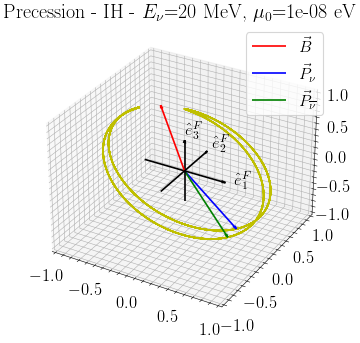}
    \caption{Snapshot of the polarization vectors corresponding to the results of figure \ref{fig:SN_Bipolar_fixed_E_and_mu_prob}, where  the left plot corresponds to the normal hierarchy and the right to the inverted (See the animations in \url{https://github.com/pedrodedin/Neutrino-Collective-Effects/blob/main/Figures/Bipolar_Oscillations_mu_constant_IH.gif} and \url{https://github.com/pedrodedin/Neutrino-Collective-Effects/blob/main/Figures/Bipolar_Oscillations_mu_constant_NH.gif}).}
    \label{fig:SN_Bipolar_fixed_E_and_mu_vec}
\end{figure}

As an example of a bipolar system, we consider a scenario with an equal amount of electron neutrinos and antineutrinos, which we call symmetric. We will also consider the neutrino-neutrino potential $\mu$ as constant during the evolution. Defining a total polarization vector for neutrinos as $\vec{P}_\nu \equiv \sum_i^N \vec{P}_{\nu_i}$ and for antineutrinos as $\vec{P}_{\overline{\nu}} \equiv \sum_i^N \vec{P}_{\overline\nu_i}$, the evolution equation can be written as follows:
\begin{equation}
\begin{split}
    \dot{\vec{P}}_{\nu} = \vec{P}_{\nu} \times \left [ \omega \vec{B} + \mu  (\vec{P}_{\nu} -\vec{P}_{\overline{\nu}}) \right],\\
    \dot{\vec{P}}_{\overline\nu} = \vec{P}_{\overline\nu} \times \left [- \omega \vec{B} + \mu  (\vec{P}_{\nu} - \vec{P}_{\overline{\nu}}) \right] .
\end{split}
\end{equation}
We can further define a sum vector $\vec{S} \equiv \vec P_\nu+ \vec P_{\overline \nu}$ and a difference vector $\vec{D} \equiv \vec P_\nu- \vec P_{\overline \nu}$ such that
\begin{equation}
\begin{split}
    \dot{\vec{S}} = \omega    \vec{D} \times  \vec{B} + \mu \vec D \times \vec  S,\\
    \dot{\vec{D}} = \omega \vec S \times  \vec B  .
\end{split}
\end{equation}
It is also useful to rewrite the sum vector as $\vec Q \equiv \vec S - \frac{\omega}{\mu} \vec{B}$, so that in the limit of a high neutrino-neutrino potential $\mu \gg \omega$ we have $\vec Q \approx \vec S $. The evolution equations then become
\begin{equation}
\begin{split}
    \dot{\vec{Q}} = \mu \vec{D} \times \vec{Q},\\
    \dot{\vec{D}} = \omega \vec Q \times  \vec B.
\end{split}
\end{equation}
As described in \citep{Hannestad2006-zw}, these are the same equations of a pendulum attracted by a force field $\vec F = \omega \vec B$ , with angular momentum $\vec L= \vec D$, length $\vec r = \vec Q$, and moment of inertia $ I = m|\vec r|^2 = \mu^{-1}$. Therefore, the system composed of $\vec{P}_\nu$ and $\vec{P}_{\overline\nu}$ oscillates around $\omega \vec{B}$, resulting in a oscillatory conversion probability \footnote{For a more detailed description check references \citep{Hannestad2006-zw,Duan:2005cp}}. It is worth noting that the sign of $\omega=\Delta m^2/2E$ depends on the neutrino mass hierarchy, such that in the normal hierarchy ($\omega>0$) there is a small amplitude of oscillation because $\vec{S}$ starts almost aligned with $\omega \vec{B}$. In the case of the inverted hierarchy ($\omega<0$), $\vec{S}$ starts anti-aligned with $\omega \vec{B}$, resulting in a complete amplitude of oscillation. This analytical description corresponds exactly to what we find in our numerical solutions shown in figures \ref{fig:SN_Bipolar_fixed_E_and_mu_prob} and \ref{fig:SN_Bipolar_fixed_E_and_mu_vec}. As expected, for the normal hierarchy, the vectors oscillate around $\vec{B}$, resulting in a small amplitude of oscillation in the survival probability, while for the inverted hierarchy, they oscillate around the opposite directions of $\vec{B}$, resulting in an almost complete conversion amplitude.

\subsubsection{Symmetric System and Varying $\mu$}

\begin{figure}[h]
    \centering
    \includegraphics[scale=0.4]{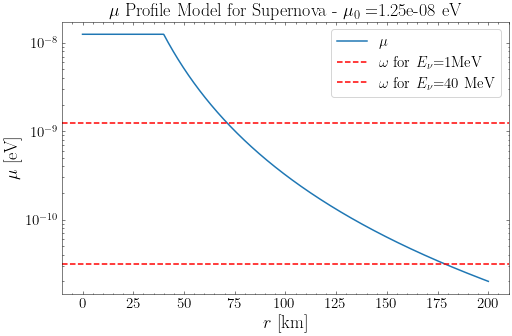}
    \caption{Profile used for the neutrino-neutrino potential $\mu$ as function of the propagated distance $r$. The horizontal dashed lines shows the vacuum frequencies $\omega$ for neutrinos with energies of 1 MeV and 40 MeV.}
    \label{fig:SN_mu_profile}
\end{figure}

A modification that we can do to the system above is to let the neutrino-neutrino potential $\mu$ change over the neutrino trajectory. As a representative of a supernova scenario we take the following profile for $\mu$ \cite{Hannestad2006-zw,Mirizzi:2015eza}, also shown in figure \ref{fig:SN_mu_profile}:
\begin{equation}
    \mu (r) =
    \left\{\begin{matrix}
     \mu_0  & r<R_\nu \\ 
    \mu_0 \left (\frac{R_\nu}{r} \right)^4 , & r>R_\nu.
    \end{matrix}\right.
\end{equation}
Here, $R_\nu$ is the neutrinosphere, for which we take a typical value of $10$ km, and $\mu_0 = 1\times 10^{-8}$ eV the initial neutrino potential. Although we are using this supernova profile for $\mu$, given that our goal is to explore supernova-like systems in the future, is worth remembering that we still working on the case of an isotropic neutrino gas.

\begin{figure}[h]
    \centering
    \includegraphics[width=0.48\textwidth]{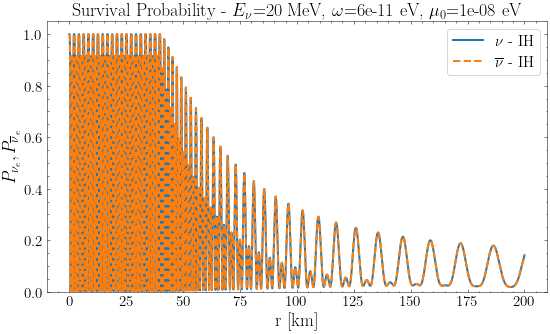}
    \includegraphics[width=0.48\textwidth]{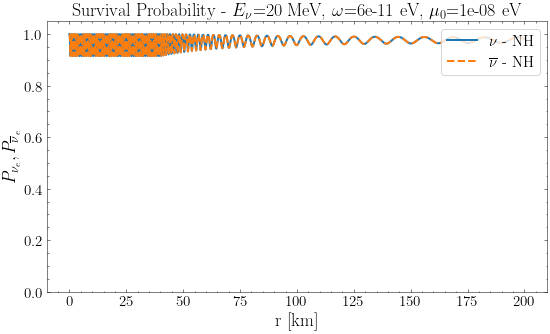}
    \caption{Survival probability for the inverted (left)  and normal (right) mass hierarchies considering the $\mu$ profile in figure \ref{fig:SN_mu_profile} and a neutrino with energy $E=20$ MeV (See the animation in \url{https://github.com/pedrodedin/Neutrino-Collective-Effects/blob/main/Figures/Bipolar_Oscillations_mu_decreasing_IH.gif} and \url{https://github.com/pedrodedin/Neutrino-Collective-Effects/blob/main/Figures/Bipolar_Oscillations_mu_decreasing_IN.gif}).}
    \label{fig:SN_Bipolar_varying_mu}
\end{figure}

As discussed in the previous section, the bipolar system behaves like a pendulum with a moment of inertia $ I = m|\vec r|^2 = \mu^{-1}$. Also, the correspondent kinetic energy $|\vec D|^2/2I$ is conserved over a period of oscillation \cite{Hannestad2006-zw}. Therefore, as $\mu$ decreases the moment inertia of the pendulum increases, and the oscillation is damped towards $\omega \vec{B}$. This damping in the oscillation will lead to an almost total conversion for the inverted mass hierarchy and practically no conversion for the normal. Figure \ref{fig:SN_Bipolar_varying_mu} shows our numerical results for this case. Again, the numerical solutions fit quite well in the analytical description. As we can see, the oscillation is damped towards total flavor conversion for the inverted hierarchy, with a small amplitude of oscillation, and no conversion for the normal hierarchy.

\subsubsection{Asymmetric System and Varying $\mu$}
\label{sec:Assym_Mono_Varying_mu}
The last scenario that we consider for our mono-energetic and isotropic gas is the one with a different initial amount of neutrinos and anti-neutrinos, i.e., an asymmetrical system. This situation corresponds to a non-vanishing initial $\vec{D}$, given by the initial difference of the lepton number. From the equations of motion, $\vec D \cdot \vec B$ is conserved, which means that the initial difference of the mass eigenstates is conserved. If the mixing angle is small ($\theta \ll 1$), this conservation is equivalent to the conservation of the net lepton number, which is the case for a supernova environment due to the mixing angle suppression by the matter potential.

\begin{figure}[h]
    \centering
    \includegraphics[width=0.49\textwidth]{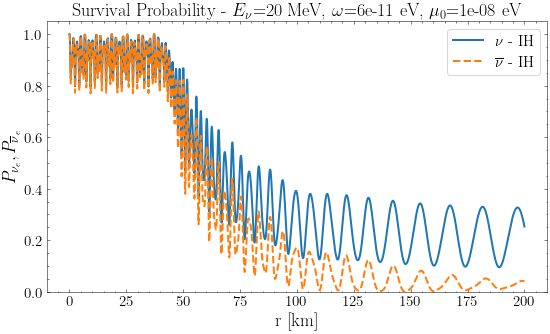}
    \includegraphics[width=0.49\textwidth]{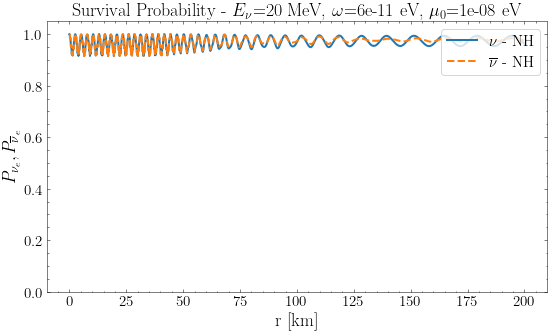}
    \caption{The same as figure \ref{fig:SN_Bipolar_varying_mu} but considering a relative flux of $1.0$ for neutrinos and $0.7$ for anti-neutrinos (See the animations in \url{https://github.com/pedrodedin/Neutrino-Collective-Effects/blob/main/Figures/Bipolar_Oscillations_Assymetric_mu_decreasing_IH.gif} and \url{https://github.com/pedrodedin/Neutrino-Collective-Effects/blob/main/Figures/Bipolar_Oscillations_Assymetric_mu_decreasing_NH.gif}).}
    \label{fig:SN_Bipolar_Assymetry}
\end{figure}

Figure \ref{fig:SN_Bipolar_Assymetry} shows the survival probability of our numerical solution for a system with a relative ratio of neutrinos to antineutrinos correspondent to $1.0:0.7$. As expected, the initial net lepton number is conserved after a period of oscillation for both mass hierarchies. We can also see the phenomenon of synchronized oscillations for $r \lesssim 100$ km. This happens when $\mu \gg \omega$ and $\vec D(0) \neq 0$ so that the precession of $\vec P_\nu$ and $\vec P_{\overline \nu}$ is dominated by $\vec D$, with precession frequencies of the order of $\mu$. Then, the vector $\vec{D}$ precesses around $\vec B$  with low oscillation frequency $\omega_{sync}= \langle\omega\rangle$, given by the average of the vacuum oscillation frequencies. In this scenario, all polarization vectors are coupled to $\vec D$ and oscillate with the same frequency $\omega_{sync}$ around $\vec B$, where the name synchronized oscillations come from.

\subsection{2\textsuperscript{nd} Case - Isotropic Neutrino Gas with Spectral Distribution}

Finally, we will explore a neutrino isotropic gas with a spectral distribution. In this case, we have equations that are similar to the mono-energetic case, with the exception that now we have one polarization vector for each momentum bin $\vec{p}_i$ and the neutrino-neutrino potential emerges as a sum over all the momenta. If we define a new total difference vector $\vec D \equiv \sum_{\vec p_2} \vec D_{\vec p_2} = \sum_{\vec p_2} (\vec{P}_{\nu,\vec{p}_2} -\vec{P}_{\overline{\nu},\vec{p}_2}) $, we may write the evolution equations as follows:
\begin{equation}
\label{eq:Precession_Spectrum}
\begin{split}
    \dot{\vec{P}}_{\nu,\vec{p}_1} = \vec{P}_{\nu,\vec{p}_1} \times \left [ \omega_{\vec{p}_1} \vec{B} + \sum_{\vec{p}_2} \mu  (\vec{P}_{\nu,\vec{p}_2} -\vec{P}_{\overline{\nu},\vec{p}_2}) \right]= \vec{P}_{\nu,\vec{p}_1} \times \left [ \omega_{\vec{p}_1} \vec{B} + \mu \vec{D} \right],\\
    \dot{\vec{P}}_{\overline\nu,\vec{p}_1} = \vec{P}_{\overline\nu,\vec{p}_1} \times \left [- \omega_{\vec{p}_1} \vec{B} + \sum_{\vec{p}_2} \mu (\vec{P}_{\nu,\vec{p}_2} -\vec{P}_{\overline{\nu},\vec{p}_2})  \right]= \vec{P}_{\overline\nu,\vec{p}_1} \times \left [- \omega_{\vec{p}_1} \vec{B} +  \mu \vec D  \right].
\end{split}
\end{equation} 
As we did for the mono-energetic case, we can write the evolution equation for the sum $\vec S_{\vec p_1}$ and difference $\vec D_{\vec p_1}$ vectors, which now depend on the momentum $\vec{p}_1$:
\begin{equation}
\begin{split}
\dot{\vec{S}}_{\vec p_1} = \omega_{\vec{p}_1} \vec{D}_{\vec p_1} \times  \vec{B} + \mu \vec D \times \vec  S_{\vec p_1},\\
\dot{\vec{D}}_{\vec p_1} = \omega_{\vec{p}_1} \vec S_{\vec p_1} \times  \vec B .
\end{split}
\end{equation}
If $\mu \gg \omega_{\vec{p}_1}$ for all modes, all $\vec{S}_{\vec p_1}$ evolve likewise, resulting in an evolution similar to the mono-energetic case for each mode \citep{Hannestad2006-zw}. However, if we consider a decreasing $\mu$, the resulting asymptotic survival probabilities will be different for each mode. As discussed in section \ref{sec:Assym_Mono_Varying_mu}, for a small mixing angle, the equations of motion lead to the conservation of the net lepton number. Therefore, if we have more neutrinos than antineutrinos and consider the inverted hierarchy scenario \footnote{In the case of the normal hierarchy we do not have any relevant conversion, as discussed for the mono-energetic case.}, we will have sufficient lepton number to completely convert the antineutrinos but only a fraction of the neutrinos, so that we have the following law for flavor conservation:
\begin{equation}
\label{eq:energy_split}
    \int_{E_{split}}^{\infty} dE \left[\phi_{\nu_e}(E) - \phi_{\nu_x}(E) \right] = \int_{0}^{\infty} dE \left[\phi_{\overline\nu_e}(E) - \phi_{\overline\nu_x}(E)\right] ,
\end{equation}
where $E_{split}$ is the energy above which total conversion occurs and none below it, leading to the phenomenon called spectral split \citep{Fogli_2007}. As we can see, this energy depends only on the initial net lepton number, which will be conserved through the propagation. For a deeper discussion on the phenomenology of the spectral split, we refer the reader to other references \citep{Raffelt2007-xu,Duan2006-dp}. Here, we will limit ourselves to using equation (\ref{eq:energy_split}) to analyze our numerical results.

%\footnote{In a nutshell, we can rewrite the equations of motion in a co-rotating frame, which do no change asymptotic probabilities. In the usual rotating frame of equation (\ref{eq:Precession_Spectrum}), for a high neutrino density $\mu \gg \omega$, the total difference vector $\vec{D}$ precesses around $\vec{B}$ with a frequency $\omega_{c}$ (synchronized oscillations). If we go to the frame where $\vec{D}$ does not rotate, each polarization vector will precess with a new vacuum frequency $\omega'=(\omega - \omega_c)$, so that as $\mu$ decreases the polarization vectors with $\omega'>0$ (low energy) will end up aligned with $\vec{B}$ and the ones with $\omega'<0$ will end up antialigned.}.  

\subsubsection{Numerical Solution}

\begin{figure}
    \centering
    \includegraphics[scale=0.45]{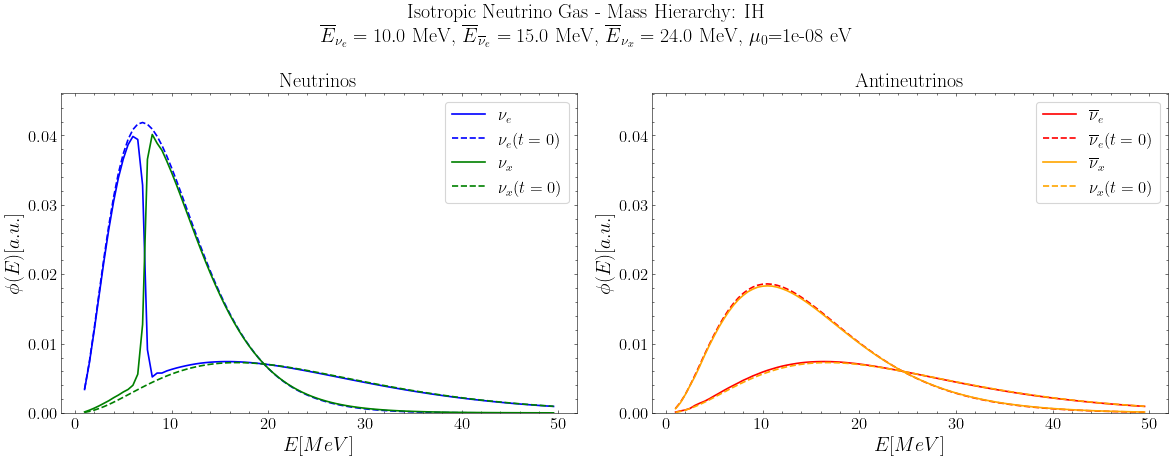}
    \caption{Initial (dashed lines) and final (solid lines) spectra for the inverted mass hierarchy, with mean energies $\overline{E}_{\nu_e}= 10$ MeV, $\overline{E}_{\overline{\nu}_e}= 15$ MeV, $\overline{E}_{\nu_x}= 24$ MeV and total energies $\varepsilon_{\nu_e}=\varepsilon_{\overline{\nu}_e}=\varepsilon_{\nu_x}$.}
    \label{fig:Isotrpic_Gas_Spectrum_IH}
\end{figure}

\begin{figure}
    \centering
    \includegraphics[scale=0.45]{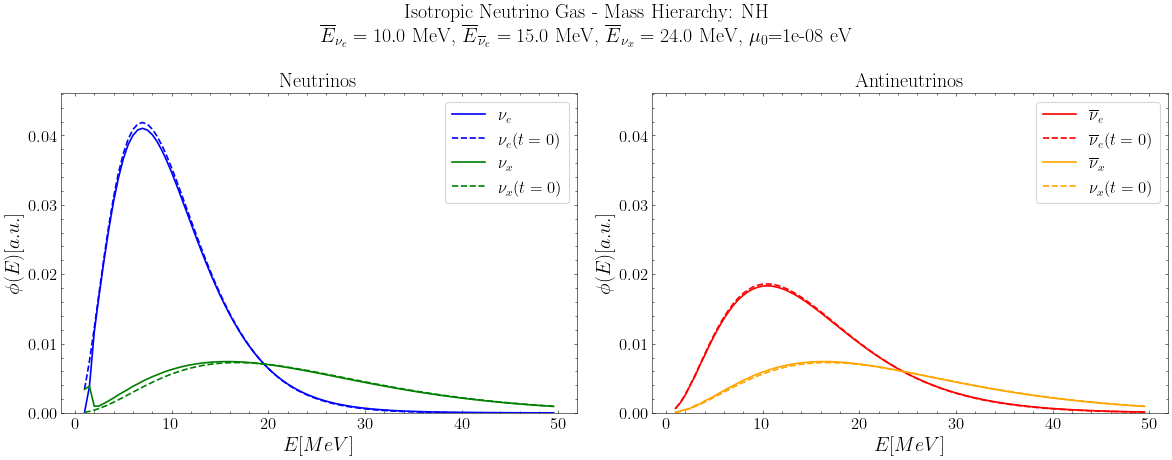}
    \caption{Initial (dashed lines) and final (solid lines) spectra for the normal mass hierarchy, with mean energies $\overline{E}_{\nu_e}= 10$ MeV, $\overline{E}_{\overline{\nu}_e}= 15$ MeV, $\overline{E}_{\nu_x}= 24$ MeV and total energies $\varepsilon_{\nu_e}=\varepsilon_{\overline{\nu}_e}=\varepsilon_{\nu_x}$.}
    \label{fig:Isotrpic_Gas_Spectrum_NH}
\end{figure}

To test this scenario in our numerical code, we decided to use the following parametrization corresponding to a supernova emission \citep{Keil_2003,Keil:2003sw}:
\begin{equation}
    f_{\alpha}(E)=N \left ( \frac{E}{\overline E}\right)^{\alpha} e^{-(\alpha+1)E/\overline E},
\end{equation}
where $\overline E$ is the averaged energy, $\alpha$ is the pinching parameter, and $N$ is a normalization constant. To implement it in the numerical code, we discretize the spectrum with an energy bin size of $\Delta E= 0.5$ MeV. Our results are shown in figures \ref{fig:Isotrpic_Gas_Spectrum_IH} and \ref{fig:Isotrpic_Gas_Spectrum_NH} for the inverted and normal mass hierarchy respectively. As we can see, for the normal hierarchy, there is no conversion at all, with the exception of some low-energy conversions \citep{Fogli:2008pt}. For the inverted hierarchy we have a total conversion in the antineutrino sector and a spectral split in the neutrino one, given that there is no sufficient lepton number to complete the conversion.

Although we are working with an isotropic neutrino gas, the spectral swap is expected to happen in some non-isotropic scenarios. For example, if we consider a supernova bulb model of emission with a single-angle approximation, the evolution equation becomes similar to the isotropic one, except for a global geometrical $D(r)$ that decreases with radius \citep{Duan2006-dp}. However, if multi-angle emission is considered, we may expect some decoherence due to the different phases along different trajectories \citep{Hannestad2006-zw}. In any case, we leave the topic of non-isotropic emissions for future work.

\section{Conclusions}
\label{sec:Conclusions}

In this paper, we have described a code to find numerical solutions for the neutrino evolution in cases where neutrino-neutrino forward scattering is relevant, which leads to the so-called collective effects. We introduced the equations to be solved and how they were implemented numerically. Then, we have shown the application of our code for different cases and compared it with other works. First, we explored the consistency of our approach by working with a solar neutrino scenario, which has shown to be in good agreement with the analytical solutions. After that, we explored the scenario of an isotropic gas of neutrinos, starting with a mono-energetic case composed of $\nu_e$ and $\overline{\nu}_e$. In this case, we explored different configurations for the relative amount of $\nu_e$ and $\overline{\nu}_e$, and different behaviors for $\mu$ (constant or decreasing). Finally, we explored a scenario with a spectral distribution for the neutrinos, where we saw the phenomena of spectral split.

Our intention with this work was not only to show our results in the effort to understand collective effects but especially to provide an open-source repository with a pedagogical description of these effects. We were motivated to do this due to the lack of open-source codes in this research field, even with a significant number of papers that rely on numerical results. In our understanding, this lack of transparency hampers reproducibility, a cornerstone of science. Therefore, we hope that this paper encourages future works to make their codes open-source or in any equivalent way to facilitate reproducibility. Besides that, we plan to continue developing the code presented here, extending it to more supernova-like scenarios. In our future work, we intend to explore non-isotropic models, such as the Bulb model \cite{Duan2006-dp}, and the relatively new phenomenon of fast oscillations \cite{Tamborra:2020cul}.

\acknowledgments
This work was supported by São Paulo Research Foundation (FAPESP) grants no. 2019/08956-2 and no. 14/19164-6. The author is thankful to Marcos V. dos Santos for useful discussions during the production of this article.

\bibliographystyle{jhep}
\bibliography{refs}

\end{document}